\documentclass[fleqn,usenatbib]{mnras}

\usepackage{newtxtext,newtxmath}

\usepackage[T1]{fontenc}
\usepackage{ae,aecompl}

\usepackage{graphicx}	
\usepackage{amsmath}	
\usepackage{comment}

\newcommand{\Msun}{~M$_{\odot}$}
\newcommand{\MMsun}{M$_{\odot}$}

\newcommand{\kms}{~km\,s$^{-1}$}

\newcommand{\HI}{H\,{\sc i}}
\newcommand{\vsys}{$v_{\rm sys}$}

\title[Discovery of galaxy merger shells in the CGM]{The Physalis system: Discovery of ORC-like radio shells around a massive pair of interacting early-type galaxies with offset X-ray emission}

\author[Koribalski et al.]{B\"arbel S. Koribalski,$^{1,2}$\thanks{E-mail: Baerbel.Koribalski@csiro.au}  Ildar Khabibullin,$^{3,4}$ Klaus Dolag,$^{3,4}$ Eugene Churazov,$^{3}$ \newauthor Ray P. Norris,$^{1,2}$ Ettore Carretti,$^{5}$ Andrew M. Hopkins,$^{6}$ Tessa Vernstrom,$^{7,8}$ \newauthor Stanislav S. Shabala,$^{9}$ Nikhel Gupta$^{8}$  \\
$^{1}$Australia Telescope National Facility, CSIRO Astronomy and Space Science, P.O. Box 76, Epping, NSW 1710, Australia \\
$^{2}$School of Science, Western Sydney University, Locked Bag 1797, Penrith, NSW 2751, Australia \\
$^{3}$Universit\"ats-Sternwarte, Fakult\"at f\"ur Physik, Ludwig-Maximilians-Universit\"at M\"unchen, Scheinerstr.~1, D-81679 M\"unchen, Germany \\
$^{4}$Max-Planck-Institut f\"ur Astrophysik, Karl-Schwarzschildstr. 1, D-85748, Garching, Germany \\
$^{5}$INAF-Istituto di Radioastronomia, Via Gobetti 101, 40129 Bologna, Italy \\
$^{6}$School of Mathematical and Physical Sciences, 12 Wally’s Walk, Macquarie University, NSW 2109, Australia \\
$^{7}$International Centre for Radio Astronomy Research, The University of Western Australia, 35 Stirling Highway, Crawley, WA 6009, Australia \\
$^{8}$CSIRO Space \& Astronomy, PO Box 1130, Bentley WA 6102, Australia \\ 
$^{9}$School of Natural Sciences, Private Bag 37, University of Tasmania, Hobart 7001, Australia 
}

\date{Accepted XXX. Received YYY; in original form ZZZ}

\pubyear{2024}

\begin{document}
\label{firstpage}
\pagerange{\pageref{firstpage}--\pageref{lastpage}}
\maketitle

\begin{abstract}
We present the discovery of large radio shells around a massive pair of interacting galaxies and extended diffuse X-ray emission within the shells. The radio data were obtained with the Australian Square Kilometer Array Pathfinder (ASKAP) in two frequency bands centred at 944~MHz and 1.4~GHz, respectively, while the X-ray data are from the XMM-Newton observatory. The host galaxy pair, which consists of the early-type galaxies ESO\,184-G042 and LEDA~418116, is part of a loose group at a distance of only 75~Mpc (redshift $z = 0.017$). The observed outer radio shells (diameter $\sim$145~kpc) and ridge-like central emission of the system, ASKAP J1914--5433 (Physalis), are likely associated with merger shocks during the formation of the central galaxy (ESO\,184-G042) and resemble the new class of odd radio circles (ORCs). This is supported by the brightest X-ray emission found offset from the centre of the Physalis system, instead centered at the less massive galaxy, LEDA~418116. The host galaxy pair is embedded in an irregular envelope of diffuse light, highlighting on-going interactions. We complement our combined radio and X-ray study with high-resolution simulations of the circumgalactic medium (CGM) around galaxy mergers from the \textit{Magneticum} project to analyse the evolutionary state of the Physalis system. We argue that ORCs / radio shells could be produced by a combination of energy release from the central AGN and subsequent lightening up in radio emission by merger shocks traveling through the CGM of these systems.
 
\end{abstract}




\section{Introduction} 
\label{sec:intro}

The "Evolutionary Map of the Universe" \citep[EMU,][]{Norris2011,Norris2021b} and the "Widefield ASKAP L-band Legacy All-sky Blind surveY" \citep[WALLABY,][]{Koribalski2012,Koribalski2020} sky surveys are the two largest science projects currently under way with the Australian Square Kilometer Array Pathfinder \citep[ASKAP,][]{Johnston2008, Hotan2021}. While both surveys produce deep radio continuum images, WALLABY's main focus is on \HI\ imaging of the nearby Universe. In addition to millions of radio continuum point sources, ASKAP surveys reveal low-surface brightness (LSB) emission structures in many shapes and sizes, including cluster halos and relics, giant radio galaxies with fading lobes, nearby star-forming disk galaxies, Galactic supernova remnants, and diffuse radio emission of yet unknown origin \citep[see, for example,][]{Koribalski2022}. With the EMU survey now $\sim$20\% complete, the source numbers in each category are rising, improving our knowledge of their properties, occurrence rates, and formation mechanisms.

\begin{figure*} 
\centering
     \includegraphics[width=17cm]{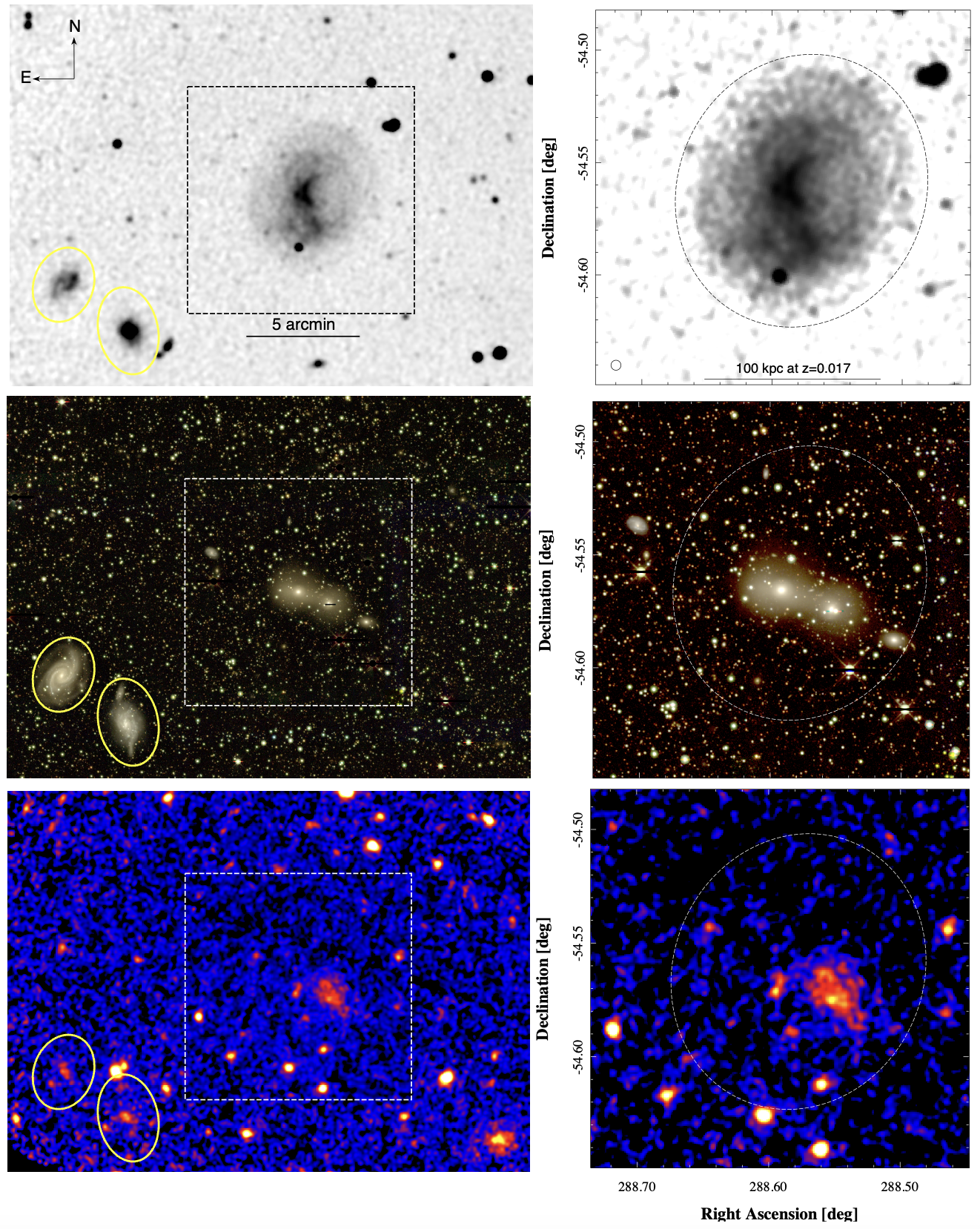}
\caption{{\bf -- Left:} The Physalis system (ASKAP J1914--5433, $z = 0.017$) and the foreground spiral galaxy pair IC~4837/9 (yellow ellipses). {\bf -- Right:} Zoom-in of the Physalis system. Its radio extent is marked by an ellipse of size $7.4' \times 6.6'$ ($PA$ = 156\degr). {\bf -- Top:} ASKAP EMU 944~MHz radio continuum image. The ASKAP synthesized beam (15\arcsec) is displayed in the bottom left corner. {\bf -- Middle:} DESI Legacy Survey DR10 $gri$-band optical image \citep{Dey2019}. {\bf -- Bottom:} XMM-Newton 0.5 -- 2.5~keV X-ray surface brightness image (resolution $\sim$ 15\arcsec). }
\label{fig:physalis-overview}
\end{figure*}

Odd radio circles \citep[ORCs,][]{Norris2021a,Koribalski2021} and similar radio rings / shells around massive early-type galaxies (see Section~1.1), are a recent addition to these extended LSB structures. Using high-resolution cosmological simulations, \citet{Dolag2023a} show that outwards moving merger shocks resembling ORCs occur in the circumgalactic medium (CGM) of massive early-type galaxies at certain stages during their formation (see Section~1.2). Such merger shells have not previously been observed around galaxy pairs or groups, although such shock fronts are known, for example, in Stephan's Quintet \citep[$z = 0.022$,][and references therein]{Geng2012,Appleton2023} and associated with the early-type galaxy pair NGC~7619/26 ($z = 0.013$) within the Pegasus group \citep{Randall2009}. \\

On larger scales ($\gtrsim$1~Mpc), merger-driven shocks are well known in galaxy clusters where they often result in arc-shaped single or double radio relics \citep[e.g.,][]{vanWeeren2019} and provide key information on the merger dynamics and evolutionary state of the cluster. Symmetric double relics are typically found in face-on cluster mergers, where they form partial shells / circles (in projection) with diameters of 2 -- 3~Mpc \citep[e.g.,][and references therein]{Koribalski2023}. In the early phase of a major merger, shocks may form in the compression zone between two fast approaching clusters \citep[see, e.g.,][]{Gu2019} and X-ray emission is typically detected around the cluster centre, between the relics. Merger shocks propagating down a steep density gradient can maintain their strength and remain efficient accelerators of particles even at large distances from the cluster center \citep[e.g.][]{Zhang2019}. \\

Here we present the Physalis\footnote{Physalis is a small orange fruit surrounded by papery shells.} system (ASKAP J1914--5433, see Fig.~\ref{fig:physalis-overview}), named after its multi-shell radio morphology. It is the closest currently known ORC-like system associated with a galaxy pair. The primary host is the S0 galaxy ESO\,184-G042 ($z = 0.017$), and its companion is LEDA~418116. See Table~\ref{tab:physalis-prop} for a summary of the Physalis system properties. In Section~1.1 we briefly describe the small sample of published ORCs (and ORC candidates) around massive early-type galaxies and summarise proposed formation scenarios. We then highlight the simulations by \citet{Dolag2023b} that suggest ORCs are formed by outwards moving merger shocks (see Section~1.2). In Section~2 we describe the ASKAP radio continuum and XMM-Newton X-ray observations and data processing, followed by our results in Section~3. We discuss our findings in Section~4, and present our conclusions in Section~5.

\begin{table} 
\centering
\begin{tabular}{lccc}
\hline
 & \multicolumn{2}{c}{Physalis system} & Ref. \\
\hline
name & \multicolumn{2}{c}{ASKAP J1914--5433}  \\
host galaxy pair & ESO\,184-G042 & LEDA~418116 \\
optical morph. type & S0 pec & SAB0: pec & RC3 \\
systemic velocity & 5135\kms & 5252\kms & W03, D12 \\
distance & \multicolumn{2}{c}{75 Mpc} \\
velocity dispersion & $176 \pm 21$\kms &  --- & W03 \\
diameter (B$_{25}$) & $80\arcsec \times 56\arcsec$  & $77\arcsec \times 44\arcsec$ & L89, M14 \\
diameter (B$_{26}$) & $136\arcsec \times 94\arcsec$  & --- & L89 \\
diameter (B$_{27}$) & $197\arcsec \times 136\arcsec$  & --- & L89 \\
position angle (B$_{26}$) & 78\degr & --- & L89 \\
B-band magnitude & 14.0 & 14.7 & L89, M14 \\
K-band magnitude & 10.0 & 10.8 & S06 \\
log stellar mass [\MMsun] & 11.1 & 10.7 & \\
\hline
\end{tabular}
\caption{Properties of the Physalis system. -- References: RC3 \citep{RC3}, W03 \citep{Wegner2003}, D12 \citep{Diaz2012}, L89 \citep{ESO-LV1989}, M14 \citep{Makarov2014}, and S06 \citep{Skrutskie2006}. }
\label{tab:physalis-prop}
\end{table}

\subsection{Odd Radio Circles}
ORCs are a newly discovered class of astronomical sources \citep{Norris2021a,Koribalski2021}, showing edge-brightened rings or shells of radio emission with diameters of $\sim$300 -- 500~kpc, but no detected counterparts at non-radio wavelengths. Here we focus on those ORCs where energetic events during the evolution of their central early-type host galaxies are likely responsible for their origins. As such galaxies contain super-massive black holes (SMBHs), their jet emission, feeding habits, etc., may play a role \citep[e.g.,][]{Velovic2023}. A number of formation scenarios have been proposed in the above papers, e.g., end-on radio lobes, giant blast waves, starburst winds. A new idea, able to explain the formation of ORCs, was proposed by \citet{Dolag2023a} who found shock structures of similar shapes and sizes around massive early-type galaxies in their simulations (see Section~1.2). Other formation scenarios include radio remnants of precessing jets seen end-on \citep{Nolting2023}, virial shocks \citep{Yamasaki2024}, AGN jet-inflated bubbles \citep{LinYang2024} and re-energised vortex rings \citep[phoenixes,][]{Shabala2024}. \\

The first three single odd radio circles -- ORC~1 \citep{Norris2021a,Norris2022}, ORC~4 \citep{Norris2021a}, and ORC~5 \citep{Koribalski2021} -- are centred on massive early-type galaxies at redshifts $z = 0.27, 0.45$ and 0.55, respectively. Using long-slit spectra, \citet{Rupke2023} recently confirm these redshifts and find that all three host galaxies have high stellar velocity dispersions ($\gtrsim$230\kms), old ($>$1~Gyr) stellar populations, and LINER-type spectra. Furthermore, [O\,{\sc iii}] imaging by \citet{Coil2024} reveals an extended (40~kpc) ionised disk around the host galaxy of ORC~4. 

Other ORCs and ORC-like sources include: SAURON, a complex, ring-like radio structure with a luminous red galaxy ($z \approx 0.55$) at its centre \citep{Lochner2023}, and ORC~J1027--4422, a partial ring with an extended central radio source \citep{Koribalski2023} as well as J0849--0457 ($z \approx 0.34$, size = 70~kpc) and J2223--4834 ($z \approx 0.27$, size = 370~kpc) found in ASKAP survey images by \citet{Gupta2022}, which appear to be associated with interacting galaxies in groups. Peculiar envelopes of diffuse radio emission around early-type galaxies were also recently highlighted by \citet{KumariPal2024a,KumariPal2024b}: J1507+3013 ($z = 0.08$, size = 68~kpc) has an ORC-like morphology and J1407+0453 ($z = 0.13$, size = 160~kpc) includes a horseshoe-shaped ring and is associated with an early-type galaxy group. \\

A few much closer systems with radio shells, which allow for the detection of hot gas via X-ray emission \citep{Kraft2022,Dolag2023b}, include the double-shell system ORC~6 ($z = 0.125$), the Cloverleaf system \citep[$z = 0.046$; Koribalski et al. 2024, in prep.;][]{Bulbul2024} -- both mentioned in \citet{Dolag2023a} -- and the Physalis system ($z = 0.017$) described here. The resolved radio structures discovered by ASKAP in the intragroup medium of these systems are shedding new light on the formation and evolutionary processes of their central host galaxies. Follow-up, high-resolution observations at a wide range of frequencies are essential to distinguish between different models. \\

For completeness we mention ORC~2 (ring size $\sim$ 80$''$ or 380~kpc) and its neighbour ORC~3 (a diffuse emission patch), discovered in ASKAP images by \citet{Norris2021a}, which are different from the single ORCs with massive central galaxies. The pair are likely the lobes of a re-started and bent radio galaxy at redshift $z \approx 0.33$ (Macgregor et al. 2024, in prep.). For a possible formation scenario see \citet{Shabala2024}. Much smaller ring-like structures are known within some radio lobes such as the Teacup \citep{Lansbury2018}, 3C\,310 \citep{Kraft2012} and the "doughnut" in NGC~6109 \citep{Rawes2018} with sizes of 10, 40 and 6~kpc, respectively, likely formed by a fast precessing jet.

\subsection{Galaxy merger shocks in simulations}

Using high-resolution cosmological simulations, \citet{Dolag2023a} find outward moving merger shocks resembling ORCs in the CGM of massive early-type galaxies, residing around the virial radius. In their simulations, shock accelerated electrons produce synchrotron emission within magnetic fields penetrating the CGM. The origin of such shocks are mergers during the formation of the central elliptical galaxy. The prediction from the simulations is that radio-detectable, outwards moving shocks / merger shells should sometimes be observable around massive early-type galaxies during their formation. Such merger shells are rare, both observationally and in the cosmological simulations. Interestingly, the simulations show hot gas residing within the shells, similar to hot halos in clusters centres bounded by radio relics. Finding such radio rings / shells around nearby galaxies, such as the Physalis system presented here, allow for much more detailed multi-wavelength studies than have been possible so far. 

\begin{table} 
\centering
\begin{tabular}{ccc}
\hline
ASKAP SB & 51574 & 51537 \\
\hline
date & 24 July 2023 & 22 July 2023\\
integration time [h] & 10 &  8 \\
field-of-view [degr$^2$] & $\sim$30 & $\sim$30 \\
centre freq [MHz] & 943.5 & 1367.5 \\
bandwidth [MHz] & 288 & 144 \\
rms [$\mu$Jy\,beam$^{-1}$] & $\sim$40 & $\sim$30 \\
resolution [arcsec] & 15 & $8.7 \times 8.0$ \\
\hline
\end{tabular}
\caption{ASKAP radio continuum observations and image properties. }
\label{tab:obs-properties}
\end{table}

\begin{table*} 
\centering
\begin{tabular}{cccccccccc}
\hline
 source name & & $z$ & size & $S_{\rm 150}$ & $P_{\rm 150}$ &  $M_{\star}$ & $M_{500}$ & $R_{500}$ & Ref. \\
  & & & [kpc] & [mJy] & [10$^{24}$ W\,Hz$^{-1}$] & [10$^{11}$~M$_\odot$] & [10$^{12}$~M$_\odot$] & [kpc] \\
\hline
 ORC~J2103--6200 & ORC~1 & 0.55 & 520 & $38 \pm 6$ & $46 \pm 7$ & $5.0 \pm 0.5$ & 260 -- 380 & 513 & 1,2 \\
 ORC J1555+2726 & ORC~4 & 0.45 & 520 & $28 \pm 3$ & $21 \pm 2$ & $1.8 \pm 0.2$ & 16 -- 25 & 234 & 1 \\
 ORC J0102--2450 & ORC~5 & 0.27 & 300 & $17 \pm 1$ & $3.9 \pm 0.2$ & $1.0 \pm 0.2$ & 2.4 -- 3.4 & 186 & 3 \\
 ORC J1027--4422 & & $\sim$0.3 & 400 & $<$17.5 & $<$5 & 0.3 & 0.51 -- 0.52 & 120 & 4 \\
 \hline
 ASKAP J1137--0050 & Cloverleaf & 0.046 & 400 & $\sim$2600 & $13.0 \pm 0.5$ & 1.4 & 7.1 -- 10, {\bf 20, 6, 9} & 323 & 5 \\
 ASKAP J1914--5433 & Physalis & 0.017 & 145 & $545 \pm 4$ & $0.37 \pm 0.01$ & 1.7 & 15 -- 20, {\bf 12, 14, 6} & 326 & here \\
\hline
\end{tabular}
\caption{Properties of ORCs and radio shell systems as well as their host galaxies. The columns are: (1--2) source names, (3) redshift $z$, (4) ORC diameter, (5) $D_{500}$ derived from $M_{500}$, (6) the measured 150~MHz flux density $S_{\rm 150}$, (7) the derived 150~MHz radio power $P_{\rm 150}$, (8) the stellar mass $M_{\star}$ of the central galaxies from \citet{Zou2019} for ORC~1, 4, 5 and derived from their $K$-band luminosity for ORC~J1027--4422, Cloverleaf and Physalis. For Physalis we give the combined mass of the host galaxy pair. (9) The $M_{500}$ estimates are based on the stellar-to-halo mass ($M_{\star} \propto M_{\rm 200c}$) relation from \citet{2020A&A...634A.135G}, where the range only reflects the uncertainty of the averaged value and does not include the scatter of individual systems. For the radio shell systems, $M_{500}$ estimates are also inferred from the X-ray temperature and luminosity using the scaling relations from \citet{2021Univ....7..139L}; the X-ray values for the Cloverleaf system are from \citet{Bulbul2024}. The last value (if present) is based on the virial mass derived from velocity data. Conversion between virial mass, $M_{\rm 200c}$ and $M_{500}$ are made based on the fitting formulae given in \citet{2021MNRAS.500.5056R}. The last column gives the references: (1) \citet{Norris2021a}, (2) \citet{Norris2022}, (3) \citet{Koribalski2021}, (4) \citet{Koribalski2023}, and (5) Koribalski et al. (2024, in prep.).}
\label{tab:orc-properties}
\end{table*}

\section{Observations and Data Analysis}
\label{sec:obs}

\subsection{ASKAP}
Fully processed radio continuum images from the Australian Square Kilometer Pathfinder (ASKAP) were obtained through the CSIRO ASKAP Data Science Archive (CASDA)\footnote{\url{https://data.csiro.au/domain/casdaObservation}}. For a description of the telescope, data processing and science highlights see \citet{Johnston2008}, \citet{Hotan2021} and \citet{Koribalski2022}. We primarily use the ASKAP 944~MHz images at 15\arcsec\ resolution as they provide the best signal to noise. ASKAP 1.4~GHz radio continuum images at $\sim$10\arcsec\ resolution are also available. A summary of the observational and image properties is given in Table~\ref{tab:obs-properties}.

\subsection{XMM-Newton}

We obtained a 21~ks long integration of the Physalis system with the X-ray space observatory XMM-Newton \citep{Jansen2001} using the European Photon Imaging Camera (EPIC) in the 0.5 -- 2.5~keV energy range. The observations were carried out on 15 Sep 2023 during Directors Discretionary Time (DDT) under Project ID 0932390101 (PI: Norbert Schartel). The data were retrieved from the XMM-Newton archive, reduced using the standard pipeline and processed via the Science Analysis System\footnote{\url{https://www.cosmos.esa.int/web/xmm-newton/sas-threads}}, as well as custom data analysis software extensively used previously for XMM-Newton data analysis of extended X-ray sources \citep[][]{2003ApJ...590..225C}. The observation was characterized by a stable particle background without significant flares, resulting in filtering out of only 1\% for the MOS 
and 10\% of the EPIC PN data. Our X-ray surface brightness images are background-subtracted as well as exposure- and vignetting-corrected (see Fig.~\ref{fig:physalis-overview}). We show the combined signal from the EPIC PN and MOS1+MOS2 detectors. The X-ray image resolution is $\sim$15\arcsec\ within the central $\sim$10\arcmin\ of the field. For the spectral analysis, we consider only the combined EPIC MOS data extracted from the region of the brightest emission, leaving further investigation of fainter diffuse emission for future exploration involving upcoming an order of magnitude deeper data (Khabibullin et al. 2024, in prep.).

\begin{figure} 
\centering
   \includegraphics[width=8cm]{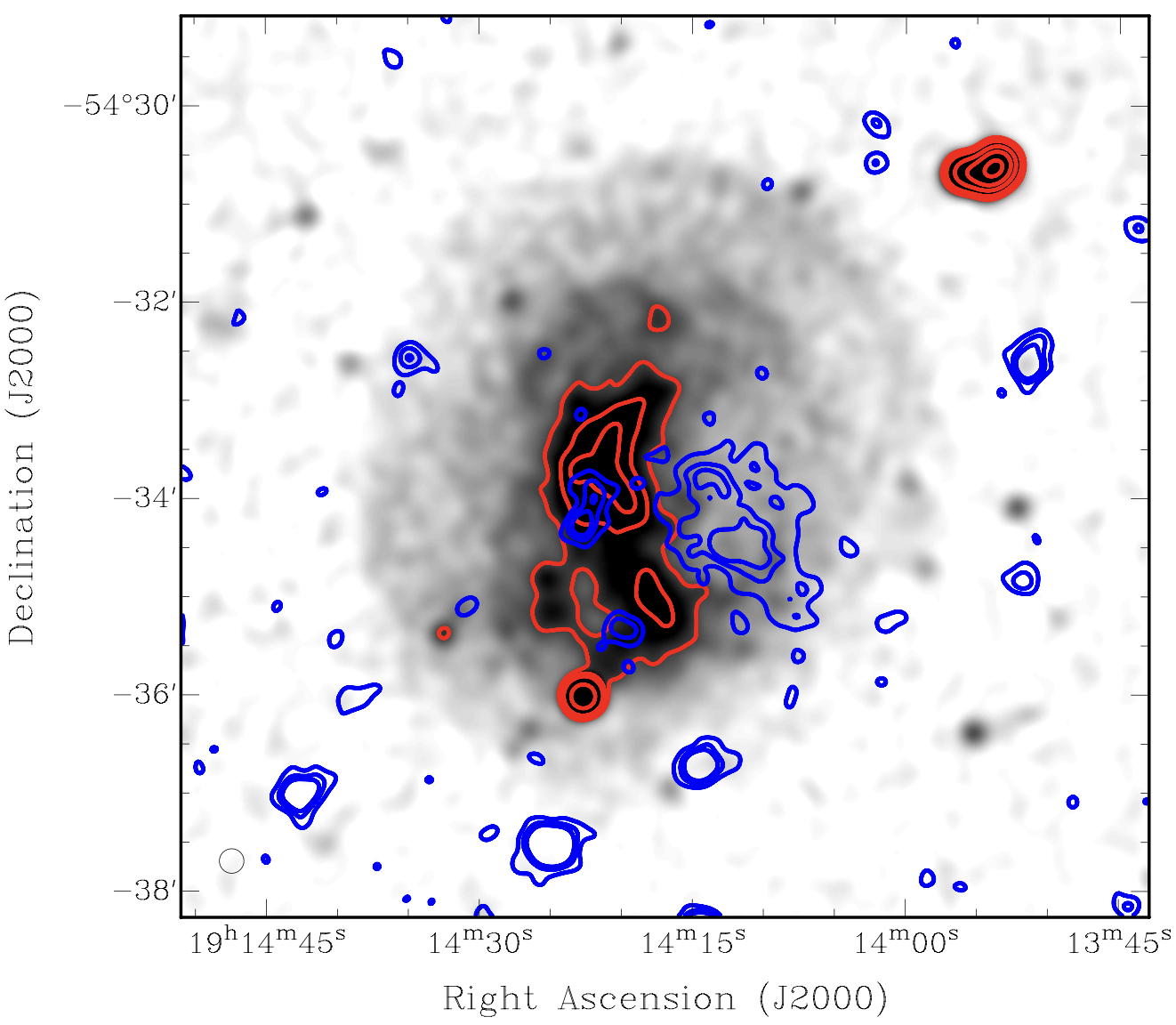}
   \includegraphics[width=8cm]{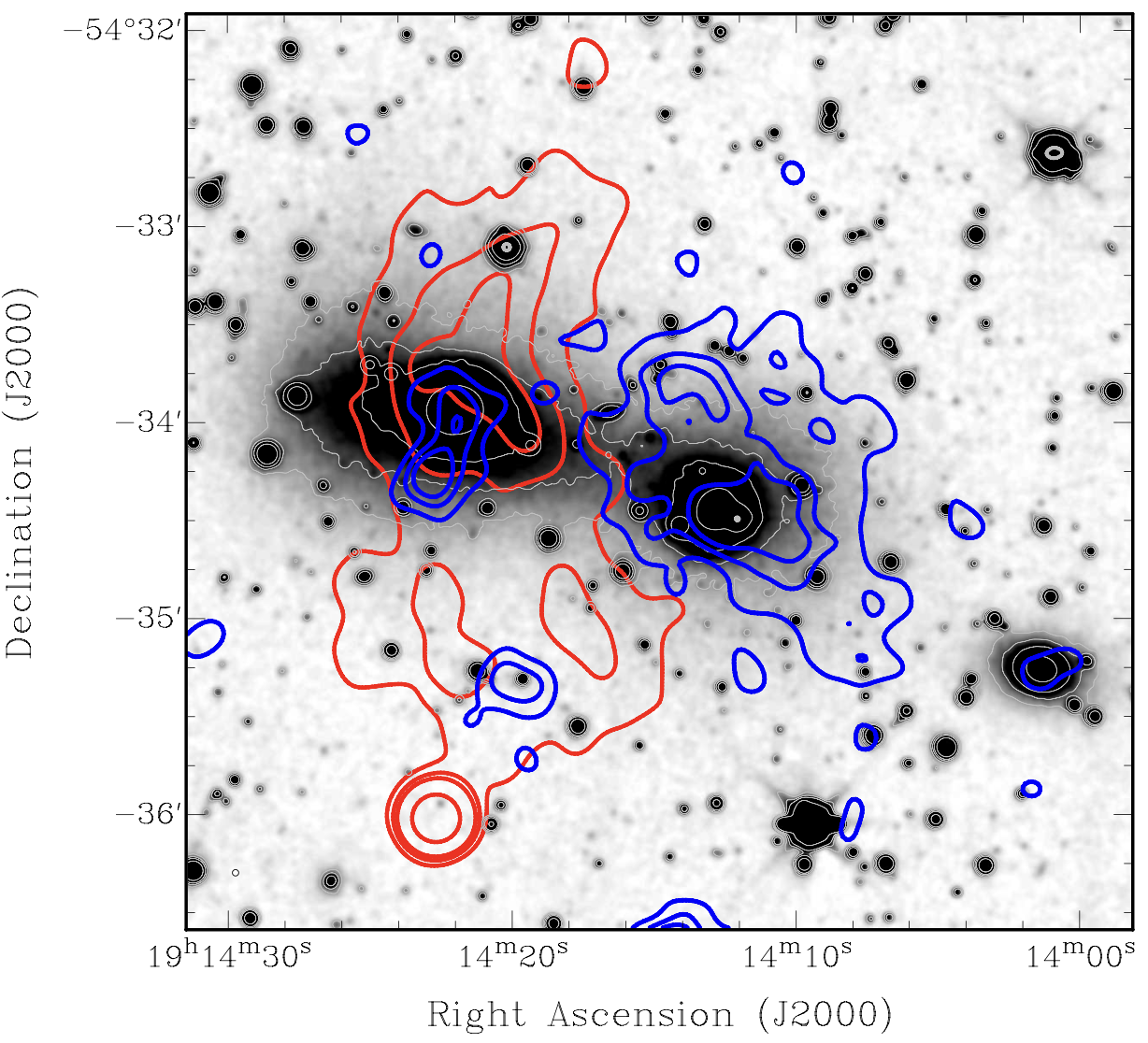}
\caption{{\bf -- Top:} Radio continuum image of the Physalis system (ASKAP J1914--5433) made by combining the ASKAP 944~MHz and 1.4~GHz images overlaid with (a) ASKAP radio continuum emission (red contours at 1, 1.5, 2, 5, 10 and 20 mJy\,beam$^{-1}$) and (b) XMM-Newton X-ray emission (blue contours at 0.0002, 0.0003 and 0.0004 counts\,s$^{-1}$\,cm$^{-2}$\,arcmin$^{-2}$). The ASKAP synthesized beam (15\arcsec) is displayed in the bottom left corner.  {\bf -- Bottom:} DESI DR10 $g$-band optical image (smoothed with a 2\arcsec\ Gaussian) of the Physalis host galaxy pair, ESO\,184-G042 and LEDA~418116, overlaid with the same contours as above. Here we highlight the offset between the bright radio ridge, centred on ESO\,184-G042, and the brightest X-ray emission centred on the neighbour, LEDA~418116. }
\label{fig:physalis-combo}
\end{figure}

\begin{figure*}
\centering
    \includegraphics[width=14cm]{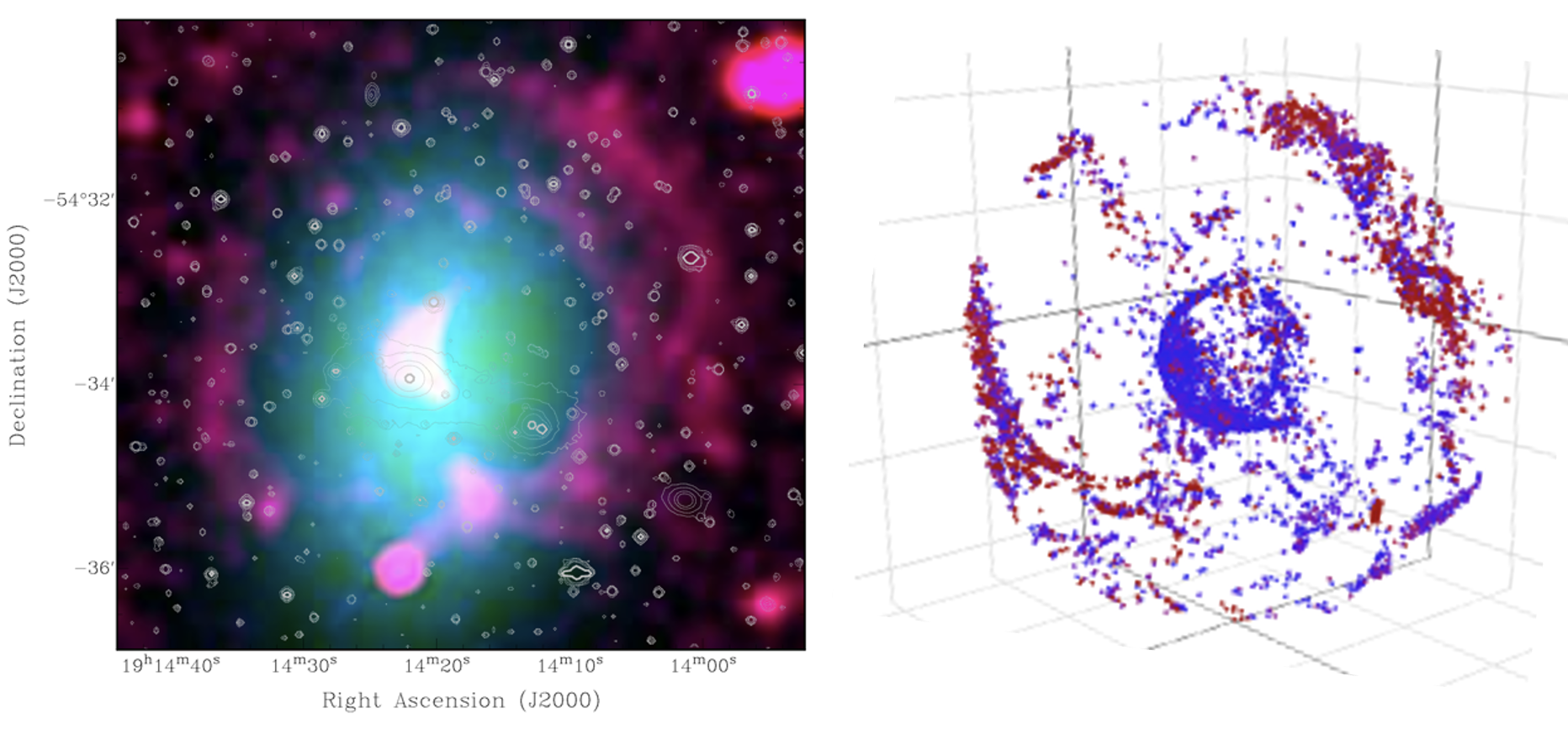}
\caption{{\bf -- Left.} The Physalis system (ASKAP J1914--5433). To emphasize the outer radio shells, we show the ASKAP 944~MHz radio continuum image (blue), a Gaussian model (green) and the residual radio emission (red) overlaid with grey contours from the optical DES-DR10 $g$-band image). {\bf -- Right.} Snapshot of a galaxy merger simulation producing large-scale relics like those seen in the left panel. Inner and outer merger structures are visible at a range of Mach numbers. Interactive 3D visualisations of the galaxy merger shocks are available at \url{http://www.magneticum.org/complements.html\#Compass} presented in \citet{Dolag2023a}.	}
\label{fig:shells+sims}
\end{figure*}

\section{Results}

To set the scene we first introduce the Physalis host galaxy system, then present our radio continuum results, followed by our X-ray results.  

\subsection{The host galaxy system}

ESO\,184-G042 is the brightest member of a loose galaxy group \citep{Tempel2018,Diaz2012}, which consists of three early-type galaxies and a small spiral galaxy that are linearly aligned (see Fig.~\ref{fig:physalis-overview} and Table~\ref{tab:physalis-prop}). The closest neighbour of ESO\,184-G042 is the galaxy LEDA~418116, located 1.4~arcmin or 30~kpc to the west. The extended envelope of diffuse stellar light around these two brightest group galaxies is remarkable, spanning at least 3.5~arcmin ($\sim$76~kpc, see Fig.~\ref{fig:physalis-combo}). The large amount of intra-group light is indicative of strong tidal interactions \citep[e.g.,][]{Spavone2018,Montes2022}. The third member of the early-type triplet is the galaxy LEDA~417985 \citep[\vsys\ = 5550\kms,][]{Jones2009}, located further west, at a projected distance of 3.3~arcmin or 72~kpc from ESO\,184-G042. The stellar masses of ESO\,184-G042, LEDA~418116 and LEDA~417985 are 1.1, 0.6 and $0.2 \times 10^{11}$\Msun\ as obtained from their $K$-band magnitudes (see Table~\ref{tab:physalis-prop}). 
Estimates of the $R_{500}$ radius and $M_{500}$ mass for the host galaxy pair are given in Table~\ref{tab:orc-properties}. The fourth group member is the spiral galaxy LEDA~418483 \citep[\vsys\ = 5143\kms,][]{Jones2009}, located 4.2\arcmin\ or 92~kpc east of ESO\,184-G042. The projected linear extent of the group is 164~kpc. Interestingly, a statistical study by \citet{Rong2024} finds alignment between loose galaxy triplets and their host filaments. The relative line-of-sight velocity of the host galaxy pair is quite small ($117 \pm 52$\kms, see Table~1), implying that the merger might be happening in the plane of the sky, as also suggested by the galaxy alignment and the morphology of the radio emission. While we detect a large envelope of diffuse intragroup light around the pair, no other signatures of tidal interactions are seen (e.g., tidal tails, shells, umbrella features). Stellar shells around elliptical galaxies have been known for a long time \citep{Malin1980,Malin1983}, and they typically form in dry (gas-poor) radial mergers \citep{Karademir2019}, while the formation of ORCs requires wet (gas-rich) and not very radial mergers (see Section~4.2). The line-of-sight velocity dispersion of the galaxies associated with the group is $\sim$180\kms, which provides a dynamical mass estimate of M$_{\rm 200c} \approx 8 \times 10^{12}$\Msun\ \citep[e.g.,][]{2008ApJ...672..122E,2021A&A...655A.115F} and $M_{500} \approx 5.8 \times 10^{12}$\Msun\ (see Table~\ref{tab:obs-properties}. This estimate is likely an underestimate, given likely merging geometry close to the picture plane.

\subsection{Radio Continuum}

Figure~\ref{fig:physalis-overview} shows the ASKAP EMU 944~MHz radio continuum image of the Physalis system and the foreground galaxy pair IC~4837/9 as well as a close-up of Physalis. The latter highlights the faint outer radio shells and bright central ridge. The radio ridge is elongated in the direction perpendicular to the axis connecting the four group galaxies. Both, shells and ridge, are centred on the massive early-type galaxy ESO\,184-G042 (see Fig.~\ref{fig:physalis-combo}). Some diffuse radio emission is also detected between the shells and the ridge. The overall size of the Physalis system is $\sim$400\arcsec\ $\times$ 320\arcsec\ ($PA \approx 156\degr$) or 145~kpc $\times$ 116~kpc. That is less than half the diameter of the more distant single ORCs \citep{Norris2021a,Koribalski2021}. The velocity dispersion of ESO\,184-G042 \citep[176\kms,][]{Wegner2003} is close to that of the ORC host galaxies \citep{Rupke2023,Coil2024}. Fig.~\ref{fig:shells+sims} highlights the distinct radio shells both in the observations (left) and in galaxy-merger simulations from the \textit{Magneticum} project (right). Red colours show the excess observed radio emission; inner and outer shells are visible. The most prominent outer radio shell is in the NW, curving towards the south, then connecting to the fainter, eastern shell, forming nearly a full circle. Secondary radio shells are seen inside the NW shell, around the bright central ridge. 

We measure total flux densities of $S_{\rm 944MHz} = 145 \pm 2$~mJy and $S_{\rm 1.4GHz} = 79 \pm 2$~mJy, resulting in a spectral index of $\alpha = -1.64 \pm 0.11$ where $S_{\nu} \propto \nu^\alpha$. For the above flux measurements a background radio source ($\sim$6~mJy), located south of the Physalis centre, was excluded. Some extended radio emission may have been filtered out, which implies that both flux estimates are lower limits.
For the brightest part of the central ridge, which is also detected in SUMSS at 843~MHz \citep[$19.1 \pm 2.4$ mJy,][]{Mauch2003}, we measure flux densities of $S_{\rm 944MHz} = 15.7$~mJy and $S_{\rm 1.4GHz} = 8.8$~mJy, resulting in $\alpha \approx -1.6$. This suggests that the outer radio shells are similarly steep.
Deeper interferometric images at a range of frequencies are needed to study the spectral index variations in the inner ridge and outer shells of Physalis.

\begin{figure*}
\begin{center}
   \includegraphics[clip,trim=2cm 5.8cm 2cm 5cm, width=0.4\textwidth]{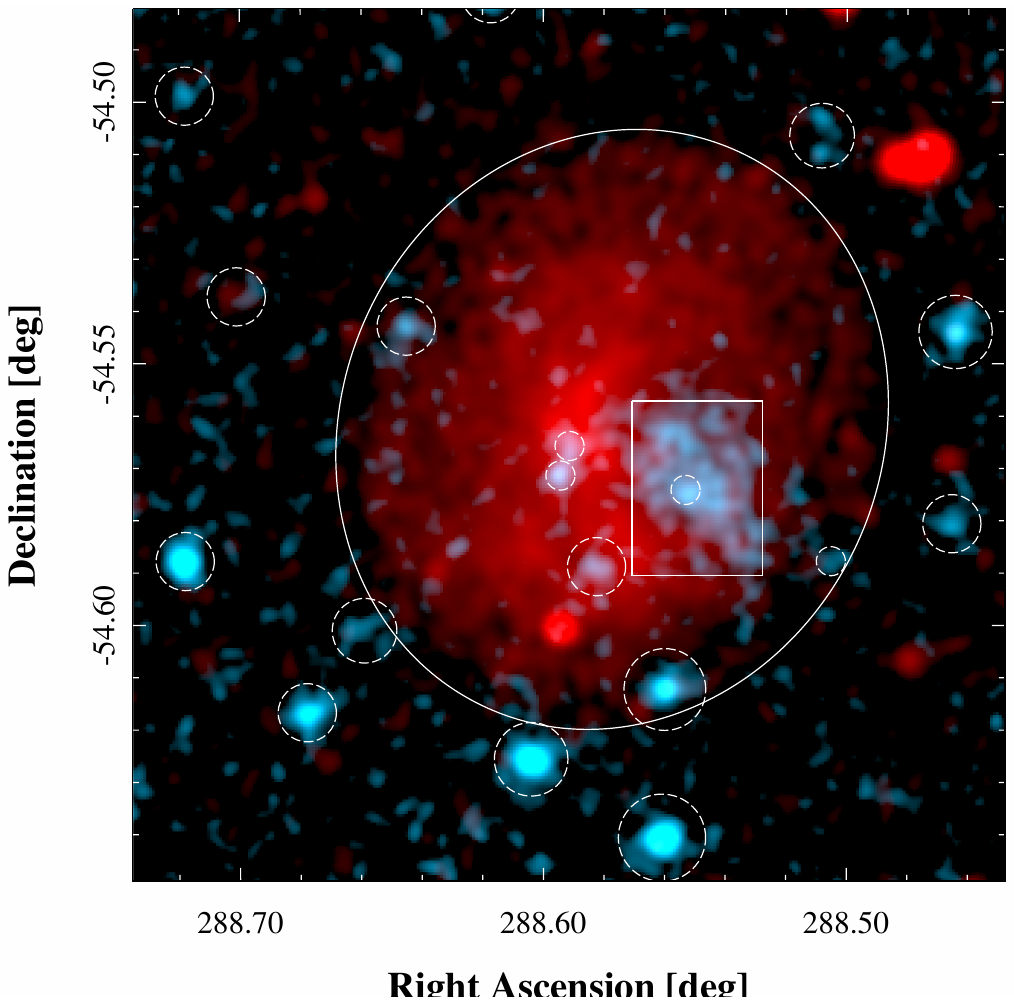}
\includegraphics[clip,trim=0cm 1.cm 2cm 0cm,width=0.59\textwidth]{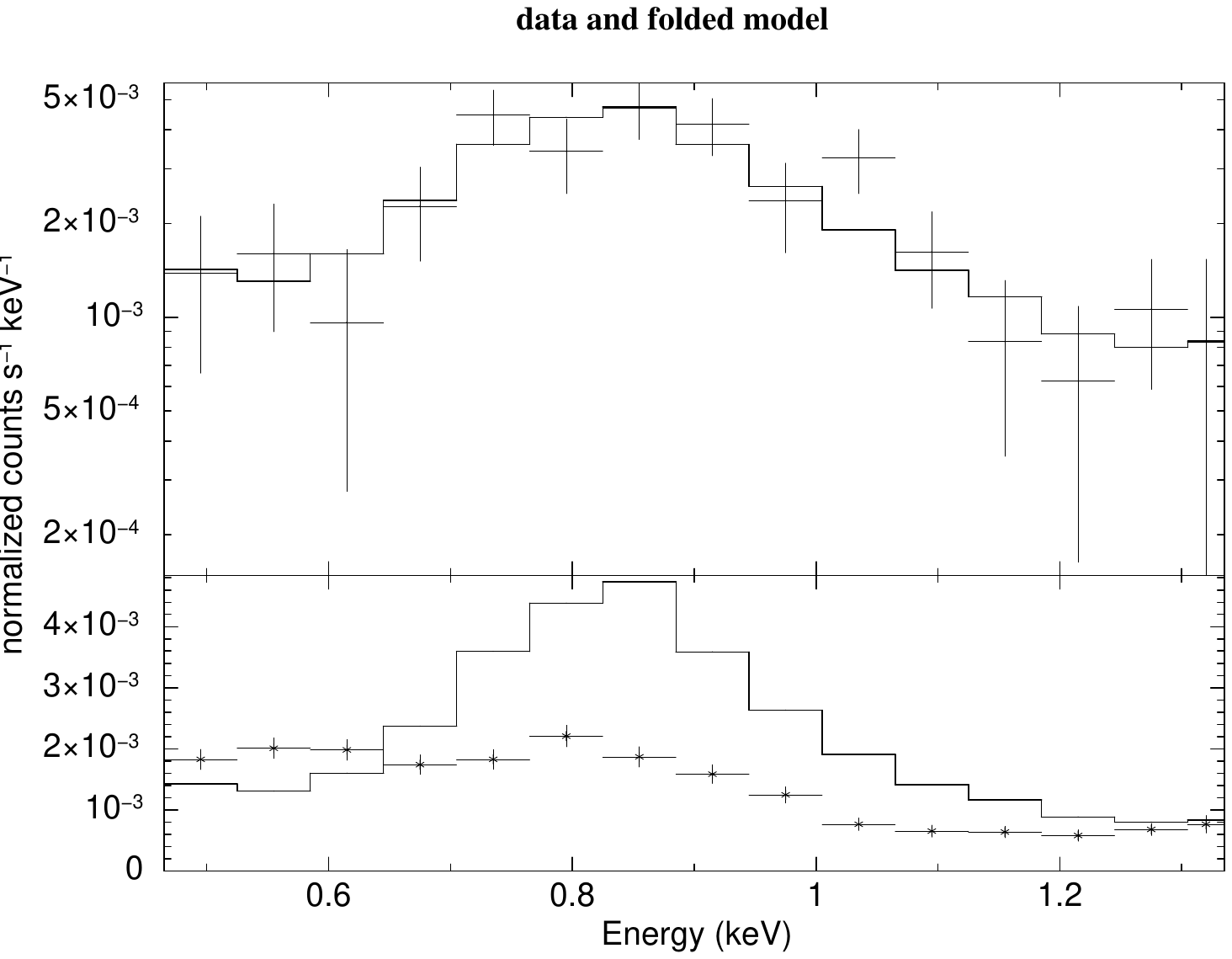}
\end{center}
\caption{Composite image of the Physalis system (left) consisting of ASKAP radio continuum emission in red and 0.5 -- 2.5~keV X-ray emission in cyan. The large ellipse highlights the extent of the radio emission (see also Fig.~\ref{fig:physalis-overview}) while the small rectangle marks the extraction region for the X-ray spectrum. Background AGN, Galactic foreground stars, and X-ray emission associated with the three early-type galaxies in the Physalis system (see Fig.~\ref{fig:physalis-combo}, bottom) are marked with small circles and were masked in the spectrum extraction procedure. The top right panel shows the X-ray spectrum from the rectangular region after subtracting the background estimate from the region confined by the ellipse, overlaid with the best fit thermal {\texttt APEC} model. The bottom right panel shows a comparison of this model with the background estimate, highlighting the energy range where the X-ray signal dominates over the background.}
\label{fig:xmm-combo}
\end{figure*}

The Physalis system is also detected at low frequencies (from 72 to 231~MHz) in the GaLactic and Extragalactic All-sky Murchison Widefield Array survey \citep[GLEAM,][]{HurleyWalker2017,HurleyWalker2022}, where it is catalogued as GLEAM~J191420--543356 with a spectral index of $\alpha = -1.0 \pm 0.1$.
Based on the catalogued source diameters, the source is extended in all GLEAM bands. For example, the catalogued flux density and source diameter in the GLEAM 171 -- 230~MHz band is $545 \pm 36$ mJy and $242\arcsec \times 183\arcsec$ ($PA = 157\degr$), respectively. The angular resolution at 200~MHz is $\sim$2\arcmin. Based on the catalogued GLEAM spectral index, we would expect ASKAP flux densities of $116 \pm 28$~mJy at 944~MHz and $80 \pm 20$~mJy at 1.4~GHz for Physalis, which agrees within the uncertainties with the measured ASKAP fluxes given above.

\subsection{X-ray Emission}

Our XMM-Newton DDT observations of the Physalis system resulted in a clear detection of diffuse X-ray emission within the radio shells. We find the hot gas to be co-spatial with the less massive galaxy (LEDA~418116) in the merging host pair and anti-correlating with the surface brightness of the radio emission (see Fig.~\ref{fig:physalis-combo}). Our 21~ks long integration already allowed for the collection of $\sim$1000 EPIC counts from the Physalis system, enabling not only the detection of extended X-ray emission but also the characterisation of its hot gas content and temperature. We confirm the absence of currently X-ray bright AGN in the host galaxy pair. 

Fig.~\ref{fig:xmm-combo} shows a composite image of the X-ray and radio continuum emission (left) as well as the extracted X-ray spectrum and best-fit emission model (right) obtained by fitting the X-ray spectrum with a thermal emission model of hot optically thin plasma. The latter is computed using {\texttt XSPEC} \citep{Arnaud1996} and an {\texttt APEC} \citep[Astrophysical Plasma Emission Code,][]{Smith2001} model with a metallicity of 0.2~$Z_{\odot}$. Assuming the line-of-sight extent of the region is comparable to its picture plane size, we estimate the electron number density of the hot X-ray gas 
at the level of $n_{\rm e} \sim 10^{-3}$ cm$^{-3}$, so the thermal pressure is $P_{\rm th} \approx 2n_{\rm e} \kappa T_{\rm e} \sim 3 \times 10^{-12}$~erg\,cm$^{-3}$. The striking anti-correlation between radio and X-ray surface brightness suggests there might be a pressure balance between X-ray and radio bright regions. In such a case, the total energy in the shell region would be $E_{\rm tot} \sim 2 \times 10^{59}$~erg, with the dense X-ray gas contributing only a few percent. The cooling time for the X-ray gas is $t_{\rm cool} \sim 4 \times 10^8$ yr, providing us with an estimate for an upper limit of the age of the observed phenomenon.

\subsection{The foreground spiral galaxy pair}

The interacting galaxy pair IC~4837/39 lies $\sim$10\arcmin\ south-east of ESO\,184-G042 (see Fig.~\ref{fig:physalis-overview}). Both galaxies are detected in the ASKAP radio and XMM-Newton X-ray images. The galaxy pair is also detected in the \HI\ Parkes All Sky Survey (HIPASS) -- catalogued as HIPASS J1915--54b -- with a flux density of $20.1 \pm 4.0$ Jy\kms\ and a systemic velocity of 2717\kms\ \citep{Koribalski2004}. We measure the following ASKAP flux densities: IC~4837 (20~mJy at 1.4~GHz, 27~mJy at 944~MHz) and IC~4839 (11~mJy at 1.4~GHz, 15~mJy at 944~MHz); the flux uncertainties are $\sim$1~mJy. Both galaxies were observed as prime candidates for the host of LIGO triggers G274296 \citep[$z_{\rm GW} > 0.42$][]{Ridley2023} and G284239 but no interesting transient candidates within the posterior constraints were identified \citep[e.g.,][]{Yang2019}. 

The interacting galaxies associated with the Physalis radio shell system ($D$ = 75~Mpc, see Fig.~\ref{fig:physalis-combo}), while not previously considered as possible hosts for these GW bursts, make interesting candidates.

\section{Discussion}

In the following we discuss the merger-driven scenario, explore a counterpart found in \textit{Magneticum}, and compare Physalis and similar systems to cluster relics.

\subsection{The merger driven scenario}

The early-type galaxy ESO~184-G042 is the brightest and most massive member of a loose group of galaxies. Its nearest neighbour, LEDA~418116, is the second brightest galaxy of the group. The distribution of the optical light in the system (see Figs.~\ref{fig:physalis-overview} \& \ref{fig:physalis-combo}) reveals an elongation of their stellar bodies as well as surrounding diffuse intragroup light, indicating on-going interactions heading towards the more violent final stage of the group's stellar core build-up. The line-of-sight velocity difference of these two galaxies is rather small, $117 \pm 52$\kms, somewhat less than the central velocity dispersion of ESO~184-G042 ($\sigma = 176 \pm 21$\kms; see Table~1) implying that either the merger proceeds in the plane of the sky or the system is currently at the apocentric phase when the line-of-sight separation between the galaxies is maximal. The former possibility is supported by the remarkable alignment of the central pair of galaxies with two other group members (LEDA~417985 and LEDA~418483), located at projected distances 70 and 90~kpc with line of sight velocity offset around 400\kms\ and $\pm$50\kms, respectively. Photometric and kinematic properties indicate that the group's total mass (dominated by dark matter) is at the level of $\sim$10$^{13}$\Msun, meaning that the observed diffuse radio emission is concentrated in the innermost 20\% of the group's $\sim$400~kpc virial radius (see also Table~\ref{tab:orc-properties}). A similar linear galaxy alignment is found for the galaxies associated with the Cloverleaf system which also exhibits ORC-like radio shells (Koribalski et al. 2024, in prep.). For a comparison of its properties with those of the Physalis system see Table~\ref{tab:orc-properties}. \\

\begin{figure}
\begin{center}
   \includegraphics[width=0.49\textwidth]{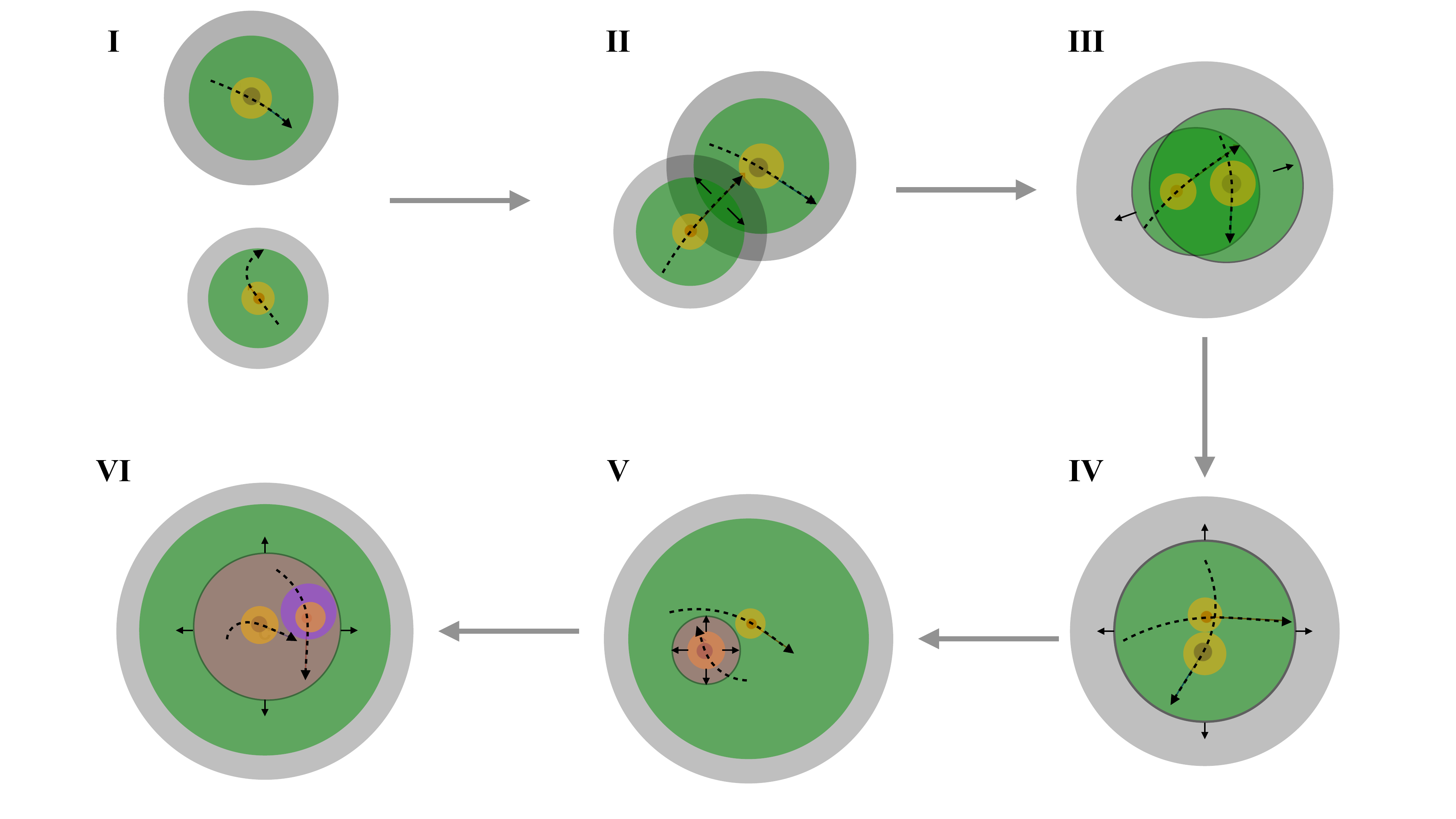}
\end{center}
\caption{Schematic description of the proposed formation scenario for Physalis ORC shown as a sequence of snapshots, $\rm I \rightarrow VI$, separated by 400~Myrs in time. The grey areas depict dark matter haloes of the two merging groups (radius $\sim R_{\rm 200c}$, i.e., the virial radius), green area -- the region where most of the hot gas content of groups -- is concentrated within $R \sim R_{500}$, the yellow regions show stellar bodies of the main group galaxies with the central regions containing respective supermassive black holes. The last two stages also contain a bubble of overheated material (shown in magenta) that was produced in an episode of powerful energy injection from the central part of one of the main galaxies, triggered by merger-induced gas inflow into it. The last snapshot shows the shocked X-ray emitting gas (in violet) that managed to survive passage of the bubble around it. The dashed arrows indicate trajectories of the main galaxies, while the solid arrows mark the direction of the shock waves launched into the system. }
\label{fig:schematic-diagram}
\end{figure}

\begin{figure} 
\centering
    \includegraphics[width=8cm]{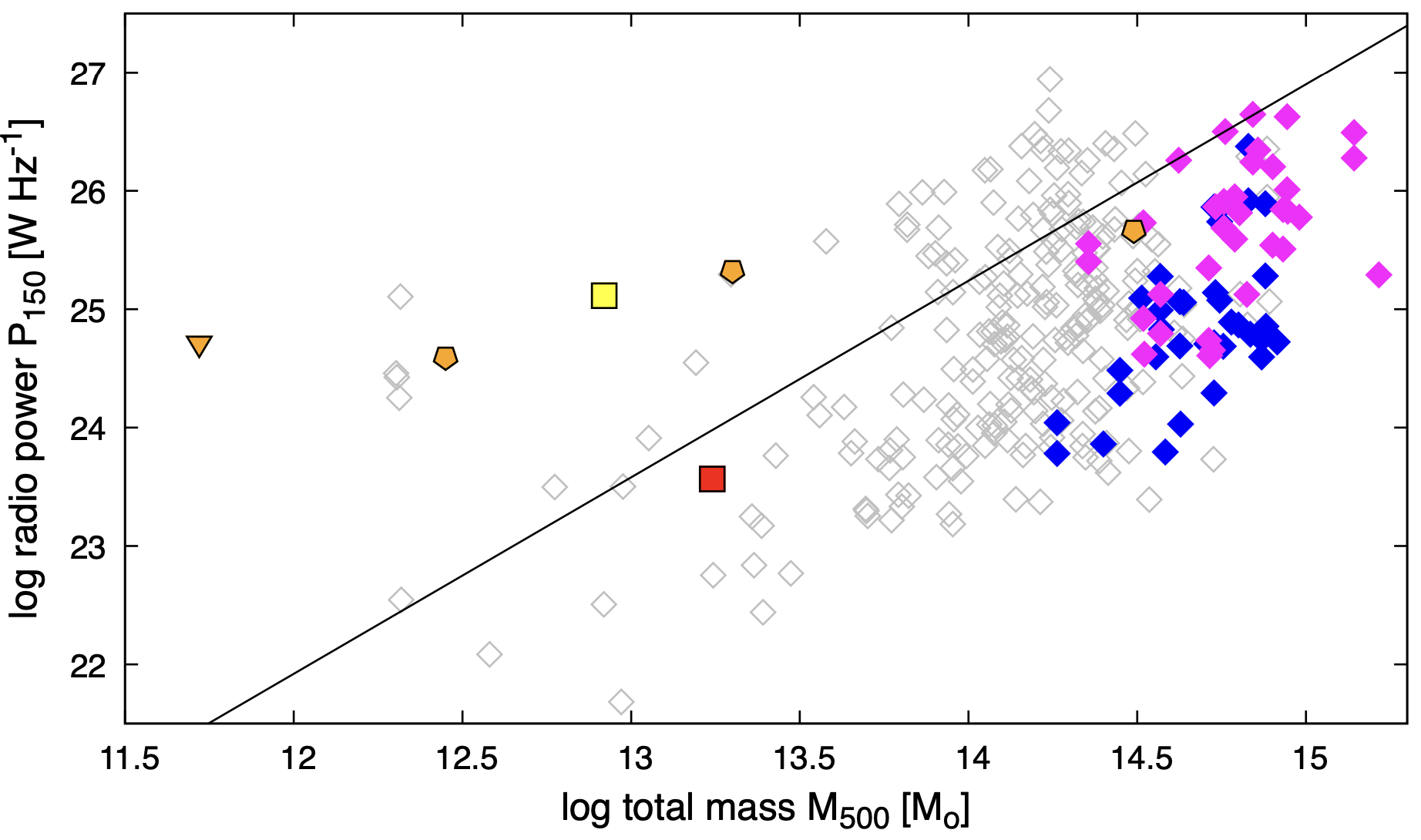}
\caption{Scaling relation of 150~MHz radio power ($P_{\rm 150}$) vs total mass ($M_{\rm 500}$) for cluster radio relics (incl. candidates) 
from \citet[][blue symbols]{Jones2023}, including double relics from \citet[][pink symbols]{deGasperin2014}, brightest cluster radio galaxies from \citet[][grey symbols]{Pasini2022} and our sample of ORCs and similar systems (see Table~\ref{tab:orc-properties}). ORCs are shown in orange, the Cloverleaf system in yellow, and the Physalis system in red. The black line indicates $P_{\rm 150} \propto M_{\rm 500}^{5/3}$ which corresponds to self-similar scaling assuming the same fraction of merging energy going into the radio power \citep[e.g.,][]{deGasperin2014}. }
\label{fig:p150-m500}
\end{figure}

The gravitational potential well of such groups is expected to be filled with diffuse X-ray emitting gas having temperatures of $\sim$0.5~keV. The XMM-Newton X-ray image of the Physalis system  reveals extended emission, $\sim$40~kpc in diameter, concentrated well within the boundaries of the radio shells. Moreover, it appears to be co-spatial with the second brightest galaxy in the group.
The measured temperature of X-ray emitting gas $\kappa T_{\rm X} \approx$ 0.7~keV is much higher than $kT \sim 0.2$~keV expected for individual elliptical galaxy of that mass. The hot gas might be a shocked remnant of the core of the second brightest group galaxy (LEDA~418116), which stands out from the more diffuse gas of the system, which is harder to detect (e.g., similar to the case of the Bullet Cluster \citealt{Markevitch2002}). It could also be the already formed core of the post-merger group, while the apparent co-spatiality with the second brightest galaxy is a transient phenomenon resulting from the merger-induced displacement of the gas with respect to the total gravitational potential of the system (shaped by distribution of merging dark matter substructures within it). Numerical simulations of the cosmological structure formation predict that such an occurrence can indeed take place, especially when gas rich substructures are infalling into a diffuse atmosphere of a large system or if the merger proceeds in a "catch up" configuration, which allows avoiding violent dispersion of the gas and causes more gentle sloshing motions instead. The schematic diagram in Fig.~\ref{fig:schematic-diagram} shows six key steps in the proposed formation scenario for the Physalis system. In Fig.~\ref{fig:p150-m500} we show the location of Physalis system and other ORCs in the radio power versus mass diagram when compared to cluster relics and radio galaxies. 

Upcoming, order-of-magnitude deeper XMM-Newton observations of the Physalis system will likely allow us to detect the full extent of the diffuse X-ray emission within the radio shells which is only hinted at in the current DDT observations, allowing for a more detailed analysis (Khabibullin et al. 2024, in prep.). \\

As noted in Section~3.2, the Physalis system is also detected in GLEAM at frequencies from 72 to 231~MHz. This allows us to obtain an independent estimate for the total energy in relativistic particles and their ages. For the observed surface brightness at 200~MHz ($545 \pm 36$~mJy) and the diameter of the Physalis system ($\sim$150~kpc), minimum energy arguments \citep[e.g.][]{BeckKrause2005} yield $B \sim$ (1--3) $\mu$G, depending on the amount of energy associated with relativistic protons, and the energy density of the non-thermal (nt) component $\sim(0.6-4) \times 10^{-13}$~erg\,cm$^{-3}$. The larger value corresponds to a commonly used assumption that there are $\sim$100 times more relativistic protons than electrons with similar energies \citep{BeckKrause2005}, which we also adopt here. The corresponding pressure of this gas phase ($1/3$ of the energy density) $P_{\rm nt} \sim 1.5 \times 10^{-13}$~erg\,cm$^{-3}$ is lower than the pressure of X-ray emitting gas  $P_{\rm th} \sim 3 \times 10^{-12}$~erg\,cm$^{-3}$. This suggests that in order to be in pressure equilibrium, either a substantial departure from the minimum energy configuration is needed or there is an extra thermal (presumably hot and low-density) component co-spatial with relativistic particles. The above arguments also provide a lower limit on the total non-thermal energy content within a sphere of radius $r = 80$~kpc: $E_{\rm nt} \sim 2 \times 10^{58}$~erg. Given that the Physalis system is detected at $\sim$1~GHz frequencies, the upper limit on the age of the latest episode of electrons' acceleration is set by their cooling time due to synchrotron and inverse Compton losses in a few $\mu$G field, $t_{\rm cool} \sim 10^8$~yr.

\subsection{A counterpart in \textit{Magneticum}}

ORCs are rare phenomena, occurring with a frequency of around one per 0.05 Gpc$^3$ as estimated by \citet{Norris2022}. This means that (1) very large simulation volumes are needed to identify counterparts, while having hosts which fall in the range of small groups, and (2) high resolution is needed to follow the evolution of this class of objects within such large volumes. One almost unique simulation which fulfills both these requirements is \textit{Box2b/hr} from the \textit{Magneticum} simulation set\footnote{\url{http://www.magneticum.org/simulations.html}}, which follows a volume of $(640 \ h^{-1}\rm cMpc)^3$ with a total of $2\cdot2880^3$ particles, allowing the particles masses to be $6.9\cdot10^8 \ h^{-1}$\Msun, $1.4\cdot10^8 \ h^{-1}$\Msun\ and $3.5\cdot10^7 \ h^{-1}$\Msun, respectively for dark matter, gas and the stellar particles, where the latter have a gravitational softening of $\epsilon = 2 \ h^{-1}\rm ckpc$ ($h = H_0/(100$\kms\,Mpc$^{-1}$) is the Hubble constant). This results in a completeness limit for galaxies of roughly $10^{9.5} \ h^{-1}$\Msun. The simulations include the treatment of cooling through tables from \citet{2009MNRAS.399..574W} as well as star formation and galactic winds with velocities of 350\kms\ \citep{2002MNRAS.333..649S}. Metal species (namely, C, Ca, O, N, Ne, Mg, S, Si, and Fe) are traces explicitly following in detail the chemical enrichment by SN-Ia, SN-II, AGB \citep{2003MNRAS.342.1025T,2007MNRAS.382.1050T} from the evolving stellar population. Black holes and associated AGN feedback are treated following \citep{2005MNRAS.361..776S}, with various improvements \citep{2010MNRAS.401.1670F,2014MNRAS.442.2304H} for the detailed treatment of the black hole sink particles and the different feedback modes. Isotropic thermal conduction of 1/20 of the standard Spitzer value \citep{2004ApJ...606L..97D} is included. The numerical scheme follows an improved formulation of Smoothed Particle Hydrodynamics (SPH) utilizing a  low viscosity scheme to track turbulence \citep{2005MNRAS.364..753D,2016MNRAS.455.2110B} and higher order SPH kernels \citep{2012MNRAS.425.1068D} for better sampling of the fluid.

\begin{figure*}
\begin{center}
   \includegraphics[width=0.8\textwidth]{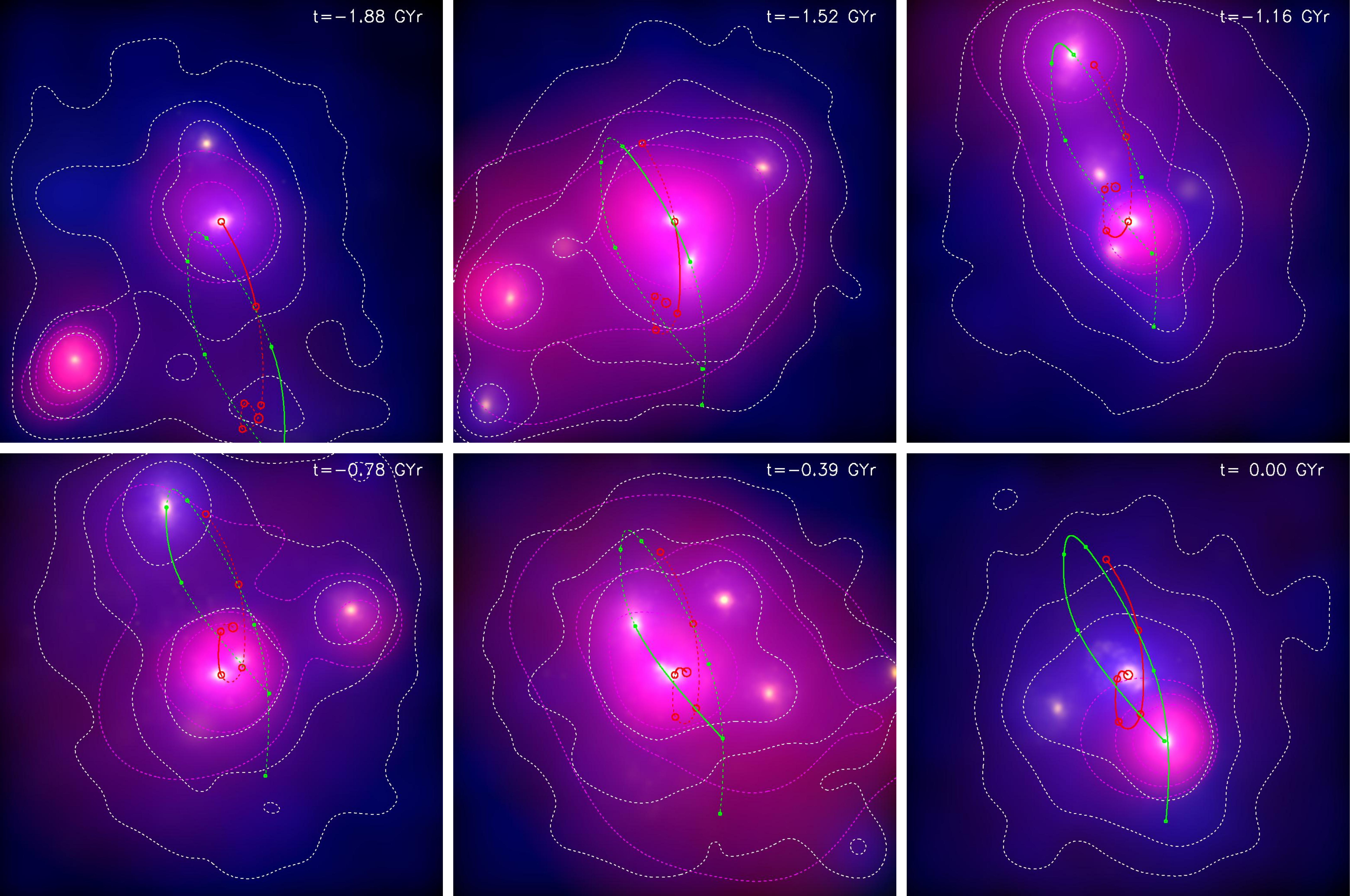}
\end{center}
\caption{Evolution (from top left to bottom right) of the candidate system selected from the {\em Magneticum Box2b/hr} simulation off the last $\approx$2~Gyr. Shown is a 200~kpc co-moving region centred on the most massive galaxy of the group. The background composite image reflects the dark matter (blue channel), the X-ray emission (red channel) and the SDSS $K$-band luminosity of the stellar component (white). To better show the assembly of the group, the white contours are drawn from the dark matter distribution, indicating that very early on the common dark matter halo is in place. In contrast, the common X-ray atmosphere (IGM, indicated as pink contours) only builds up gradually reaching the most similar shape compared to the dark matter roughly five hundred million year before the final time at which the group was selected as counterpart. The circles along the trajectories of the two main galaxies are indicating the size of the black holes in the centres of the galaxies. The black hole in he most massive galaxy grows significantly between the last times shown, releasing large amount of energy in form of feedback, which energizes and lifts a large fraction of the IGM to larger distances, so that only the part associated to the second galaxy stays visible. }
\label{fig:mag_candidate}
\end{figure*}

The cosmology adopted for these simulations is the WMAP7 from \citet{2011ApJS..192...18K}, with a total matter density of $\Omega_{\rm m} = 0.272$, of which 16.8\% are baryons, the cosmological constant $\rm \Lambda_{0} = 0.728$, the Hubble constant $\rm H_{0} = 70.4$\kms\,Mpc$^{-1}$, the index of the primordial power spectrum $\rm n=0.963$ and the overall normalisation of the power spectrum $\sigma_{8} = 0.809$. Halos and galaxies are identified using \textsc{SubFind} \citep{2001MNRAS.328..726S, 2009MNRAS.399..497D}, where the centre of a halo is defined as the position of the particle with the minimum of the gravitational potential. The virial mass, $M_{\rm vir}$ is defined through the spherical overdensity as predicted by the generalised spherical top-hat collapse model \citep{1996MNRAS.282..263E} and, in particular, it is referred to $R_{\rm vir}$, whose overdensity to the critical density follows Eq.~6 of \citet{Bryan98}.

Previous studies have demonstrated that the galaxy physics implemented in the \textit{Magneticum} simulations leads to an overall successful reproduction of the basic galaxy properties, like the stellar mass-function \citep{2017ARA&A..55...59N,2022arXiv220109068L}, the environmental impact on galaxy properties \citep{2019MNRAS.488.5370L} as well as the associated AGN population and their evolution \citep{2014MNRAS.442.2304H,2016MNRAS.458.1013S,2018MNRAS.481.2213B}. At cluster scales, the \textit{Magneticum} simulations have demonstrated to reproduce the observable X-ray luminosity-relation \citep{2013MNRAS.428.1395B}, the pressure profile of the ICM \citep{2017MNRAS.469.3069G} and the chemical composition \citep{2017Galax...5...35D,2018SSRv..214..123B} of the ICM, the high concentration observed in fossil groups \citep{2019MNRAS.486.4001R},  as well as the gas properties in between galaxy clusters and groups \citep{Biffi22,2023A&A...675A.188A}. On larger scales, the \textit{Magneticum} simulations demonstrated to reproduce the observed SZ-Power spectrum \citep{2016MNRAS.463.1797D} as well as the observed thermal history of the Universe \citep{2021PhRvD.104h3538Y,2024PhRvD.109f3513C}. Most importantly for this work, the \textit{Magneticum} simulations reproduces the observed level of entropy of the gas at group scales much better than other, current, cosmological hydrodynamical simulations, indicating that the model injects a realistic amount of feedback energy also at group scales \citep{2024arXiv240117276B}. Furthermore, the simulations reproduce X-ray scaling relations for galaxy groups as observed by eROSITA, where the central entropy of the gas as regulated by the AGN feedback plays a key for the detectability of the groups based on their X-ray luminosity (Marini et al., in prep). 

\begin{figure*}
\begin{center}
    \includegraphics[width=0.8\textwidth]{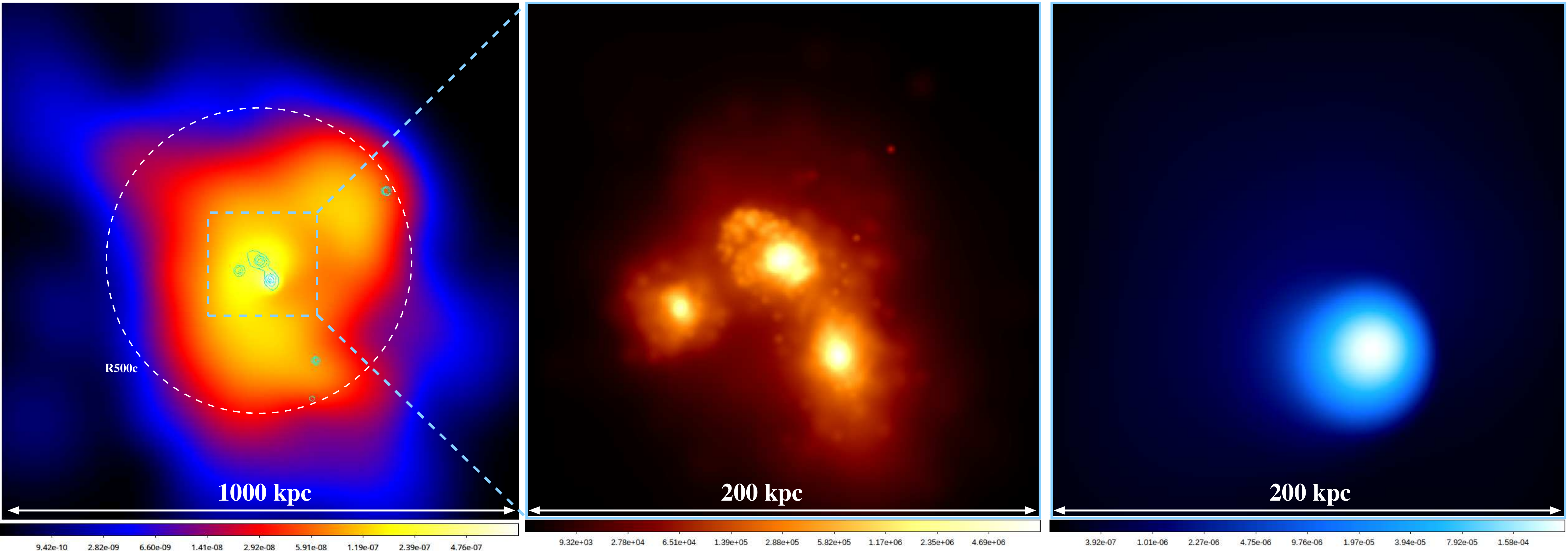}
\end{center}
\caption{Mock images from the simulated counterpart of the Physalis system. {\bf -- Left:} The large-scale thermal Sunyaev Zeldovich signal, highlighting the pressure distribution within the halo and the impact of the AGN feedback. Overlaid in cyan are the contours from the SDSS $r$-band mocks, indicating the positions of the central galaxies. The white, dashed circle indicates the halo radius of $R_{\rm 500c}$. {\bf -- Middle:} The optical SDSS $r$-band mocks in the central 200~kpc region. {\bf -- Right:} The X-ray surface brightness in the same region based on the emission in the 0.1 -- 2.4~keV band, highlighting the offset between the X-ray emission peak, which is centered on the companion galaxy and the central galaxy. }
\label{fig:mag_prop}
\end{figure*}

To find a counterpart to the Physalis system within {\it Box2b/hr} we match pairs of galaxies in the simulation based on the observational properties. After identifying such systems we can study their formation history and derive hints on possible formation scenarios as well as estimating the occurrence rate of ORCs resulting from outwards moving merger shocks. The main selection criteria from the observation was the close proximity of the two galaxies, their stellar mass as well as the remaining of some gas centered around the smaller of the two galaxies. Given the close distance of the two galaxies and their stellar masses, the dark matter halos of the two galaxies, being typically $\approx 10$ times more extended than the stellar body, will be already identified as one, large scale over-dense region. Assuming virial masses ($M_{\rm vir}$) in the range $1-3 \times 10^{13}$\Msun, we found a starting set of $\sim$26000 low mass groups within the volume of the simulation (e.g., 0.75 Gpc$^3$). We then applied the following, additional requirements: (a) The system contains two massive galaxies, where the second most massive has at least $1/3$ of the total stellar mass; (b) The 3D distance between the two massive galaxies is less than 70~kpc; (c) The galaxies' stellar masses within a 20~kpc aperture differ by more than 10\%; (d) At least one galaxy has a total gas mass greater than $10^{10}$\Msun\ within the 20~kpc aperture; (e) The smaller galaxy (in terms of stellar mass) has at least 10\% more hot gas ($T > 10^6$~K) than the larger galaxy within the 20~kpc aperture. These additional requirements left us with only 10 halos within the simulation volume.

Forcing the galaxies not to be star forming and not showing obvious tidal features, we are left with only one candidate system, for which we studied the evolution in detail. This already demonstrates that such merger configurations are quite rare, being consistent with the low number of observed ORCs. 
A visual impression of its evolution is shown in Fig.~\ref{fig:mag_candidate}. The detailed investigation of the evolution of this peculiar system showed several interesting details. On the one hand the group was generally X-ray bright before the first passage of the two galaxies. However, during this first passage, cold gas is channeled onto the BH of the more massive galaxy which sits closer to the center of the total potential and the common gas envelope is energized and expanded so that it is below the direct detection threshold of our current XMM observations. A remaining fraction of concentrated, hot gas is still bound to the local potential of the smaller, satellite galaxy and matches in its properties the observations. Interestingly, the BH in the more massive galaxy grows by $\sim$10$^8$\Msun\ during the first passage, corresponds to an energy release of $\sim$10$^{59}$ erg, which matches the energy content inside the radio emitting region of the ORC inferred by the X-ray observations. Fig.~\ref{fig:mag_prop} shows the mock observations of the simulated system at the time when it was selected in the simulation, where the synthetic Sunyaev Zeldovich (SZ) map also shows the distribution of the pressure within the system. This broadly confirms the picture that such ORCs / radio shells could be produced by a combination of energy release from the AGN and subsequent lightening up in radio by merger shocks traveling through the CGM of these systems. \\

In Fig.~\ref{fig:p150-m500} we highlight the location of the Physalis system (red square) and other ORCs in the radio power vs mass diagram typically used to explore cluster scaling relations. As more ORCs are discovered and used to populate this diagram, we can explore their similarity to cluster relics. While the latter have additional physical processes, which makes the relation steeper, ORCs could indicate that they are as efficient in converting the merger energy into radio power than the most efficient clusters, or even more (in the case of the Cloverleaf), indicating that energization by the AGN seems to be a quite efficient process.

\section{Conclusions}
 
The ASKAP discovery of the ORC-like Physalis system together with the XMM-Newton detection of diffuse X-ray emission is a great opportunity to investigate and understand the formation mechanism of such large-scale radio shells around such a nearby, interacting galaxy pair. While the shells are centred on the brightest and most massive group galaxy, ESO\,184-G042, the hot gas is found offset by 30~kpc in and around the less-massive companion galaxy, LEDA~418116 (see Fig.~\ref{fig:physalis-combo}). Such offset between the central radio emission and the hot gas, also just recently seen in the Cloverleaf system \citep{Bulbul2024}, is likely a signature of the unsettled state of the system, similarly to what is sometimes observed in clusters \citep[e.g.,][]{Rosignoli2024}. The orientation of the radio shells (perpendicular to the linearly aligned group galaxies) together with the extended envelope of intragroup light surrounding the central galaxy pair, and the offset hot gas, suggests a full merger is in progress. We suggest that the Physalis radio shells are formed by outwards moving merging shocks \citep{Dolag2023a}, and while resembling cluster radio relics \citep[see also][]{Koribalski2023}, they are smaller and associated with mergers in galaxy groups. An investigation of the radio power to mass relation is under way. 

In the Physalis system, the derived temperature and density of the hot X-ray emitting gas combined with the ASKAP flux densities allows us to place constrains on the energetics of the event that led to the ORC / radio shell formation and differentiate between possible scenarios. The required release of 10$^{59}$~erg over a period of a few hundred million years might be provided by an episode of SMBH activity triggered by gas inflows during the merger of two central galaxies accompanying the build-up of a compact group. Strong shocks are needed to accelerate electrons from the thermal pool, but even weak shocks may be able to re-accelerate fossil electrons. We were able to identify and explore one matching galaxy system in the \textit{Magneticum} simulations and present snapshots of its evolution, underlining the rarity of such energetic events. This is an exciting field which is growing rapidly as more ORCs and radio shell systems are being discovered.

\vspace{-0.3cm}

\section*{Acknowledgements}

We thank Norbert Schartel for his support of the XMM-Newton DDT observations.  IK and KD are supported by the Excellence Cluster ORIGINS which is funded by the Deutsche Forschungsgemeinschaft (DFG, German Research Foundation) under Germany's Excellence Strategy -- EXC-2094-390783311 and the COMPLEX project from the European Research Council (ERC) under the European Union’s Horizon 2020 research and innovation program grant agreement ERC-2019-AdG 882679. \\

The Australian SKA Pathfinder is part of the Australia Telescope National Facility (ATNF) which is managed by CSIRO. Operation of ASKAP is funded by the Australian Government with support from the National Collaborative Research Infrastructure Strategy. ASKAP uses the resources of the Pawsey Supercomputing Centre. Establishment of ASKAP, the Murchison Radio-astronomy Observatory (MRO) and the Pawsey Supercomputing Centre are initiatives of the Australian Government, with support from the Government of Western Australia and the Science and Industry Endowment Fund. This paper includes archived data obtained through the CSIRO ASKAP Science Data Archive (CASDA). We acknowledge the Wajarri Yamatji as the traditional owners of the Observatory site. \\

This project used public archival data from the Dark Energy Survey (DES). Funding for the DES Projects has been provided by the U.S. Department of Energy, the U.S. National Science Foundation, the Ministry of Science and Education of Spain, the Science and Technology Facilities Council of the United Kingdom, the Higher Education Funding Council for England, the National Center for Supercomputing Applications at the University of Illinois at Urbana-Champaign, the Kavli Institute of Cosmological Physics at the University of Chicago, the Center for Cosmology and Astro-Particle Physics at the Ohio State University, the Mitchell Institute for Fundamental Physics and Astronomy at Texas A\&M University, Financiadora de Estudos e Projetos, Funda{\c c}{\~a}o Carlos Chagas Filho de Amparo {\`a} Pesquisa do Estado do Rio de Janeiro, Conselho Nacional de Desenvolvimento Cient{\'i}fico e Tecnol{\'o}gico and the Minist{\'e}rio da Ci{\^e}ncia, Tecnologia e Inova{\c c}{\~a}o, the Deutsche Forschungsgemeinschaft, and the Collaborating Institutions in the Dark Energy Survey.

\vspace{-0.3cm}

\section*{Data availability} 

The ASKAP data products are publicly available in the CSIRO ASKAP Science Data Archive (CASDA) at \texttt{data.csiro.au/domain/casdaObservation}. And the XMM-Newton data are available at \texttt{www.cosmos.esa.int/web/xmm-newton/too-details}. \\

Additional data processing and analysis was conducted using the {\sc miriad} software\footnote{https://www.atnf.csiro.au/computing/software/miriad/} and the Karma visualisation\footnote{https://www.atnf.csiro.au/computing/software/karma/} packages.







\bibliographystyle{mnras}
\bibliography{mnras,biblio} 





\appendix


\bsp	
\label{lastpage}
\end{document}